\documentclass[10pt,conference]{IEEEtran}

\usepackage{cite}

\ifCLASSINFOpdf
   \usepackage[pdftex]{graphicx}
   \graphicspath{{figs/}}
   \DeclareGraphicsExtensions{.pdf,.jpeg,.png}
\else
   \usepackage[dvips]{graphicx}
   \graphicspath{{../figs/}}
   \DeclareGraphicsExtensions{.eps}
\fi

\usepackage[cmex10]{amsmath}

\usepackage{amsthm}

\usepackage{algorithmic}
\usepackage{algorithm}

\usepackage{array}

\ifCLASSOPTIONcompsoc
  \usepackage[caption=false,font=normalsize,labelfont=sf,textfont=sf]{subfig}
\else
  \usepackage[caption=false,font=footnotesize]{subfig}
\fi

\usepackage{url}
\usepackage{xspace}
\usepackage{xcolor}
\usepackage{tikz}
\usepackage{multirow}
\usepackage{booktabs}
\usepackage{array}
\usepackage{tabularx}
\usepackage[english]{babel}
\usepackage{hyperref}
\usepackage{amssymb}

\usepackage[table]{xcolor}
\definecolor{rowgray}{gray}{0.94}

\addto\extrasenglish{%
}

\newcommand{\appref}[1]{\hyperref[#1]{App.~\ref*{#1}}}

\usepackage{listings}
\definecolor{codebg}{RGB}{245,245,245}
\definecolor{coderule}{RGB}{210,210,210}
\newif\iffinal
\finaltrue
\newcommand{\cmtid}{170}

\iffinal
\else
\usepackage[switch]{lineno}
\fi

\newcommand{\systemname}{\textsc{ContraVis}\xspace}
\newcommand{\myparagraph}[1]{\smallskip\noindent\textbf{#1}\xspace}
\newcommand{\eg}{\emph{e.g.}\xspace}

\begin{document}

\title{\systemname: Evidence-Grounded Visual Analytics for Contradiction Review in Legal Contracts}

\iffinal

\author{\IEEEauthorblockN{%
Luis Sante,
Paula Lima,
Mariana Rocha, and
Jorge Poco}
\IEEEauthorblockA{%
School of Applied Mathematics, Funda\c{c}\~{a}o Getulio Vargas (FGV)\\
\{taipe.luis, paula.lima.1, mariana.rocha\}@fgv.edu.br, jorge.poco@fgv.br}}

\else
  \author{SIBGRAPI Paper ID: \cmtid \\ }
  \linenumbers
\fi

\maketitle

\begin{abstract}
Legal contracts are structurally complex documents in which contradictions may emerge across distant and interconnected provisions.
Although large language models (LLMs) improve legal language understanding, contradiction analysis remains a human-centered and evidence-grounded review task.
We present \systemname, a visual analytics system for human-in-the-loop contradiction analysis in legal contracts.
The system models contracts as typed paragraph graphs that combine explicit contractual references with semantic relationships between paragraphs.
This graph plays a dual role: it conditions LLM reasoning and serves as the interactive representation the analyst explores, keeping model context and human inspection aligned across coordinated views.
In a controlled comparison, graph-conditioned reasoning recovered more injected contradictions than standalone LLM analysis as contract length grew, while surfacing additional candidates for analyst validation.
A formative study with contract-domain lawyers indicated that in-context evidence comparison supported contradiction validation, and we distill design implications for evidence-grounded, LLM-assisted document review.
\end{abstract}

\IEEEpeerreviewmaketitle

\newcommand{\figworkflow}{
\begin{figure}[t]
    \centering
    \includegraphics[width=\columnwidth]{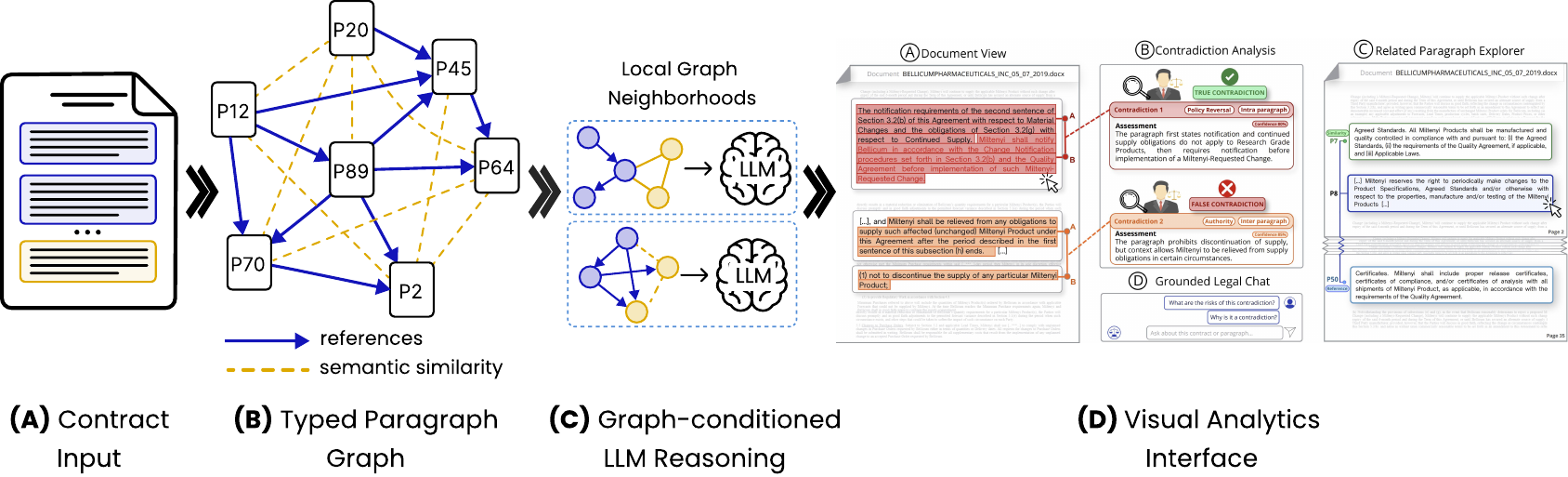}
    \caption{Human-in-the-loop workflow: from a contract (A), \systemname builds a typed paragraph graph (B), performs graph-conditioned LLM reasoning to surface candidates with supporting evidence (C), which experts validate through coordinated views (D).}
    \label{fig:workflow}
\end{figure}
}

\newcommand{\uiinterface}{
\begin{figure*}[t]
  \centering
  \includegraphics[width=0.9\linewidth]{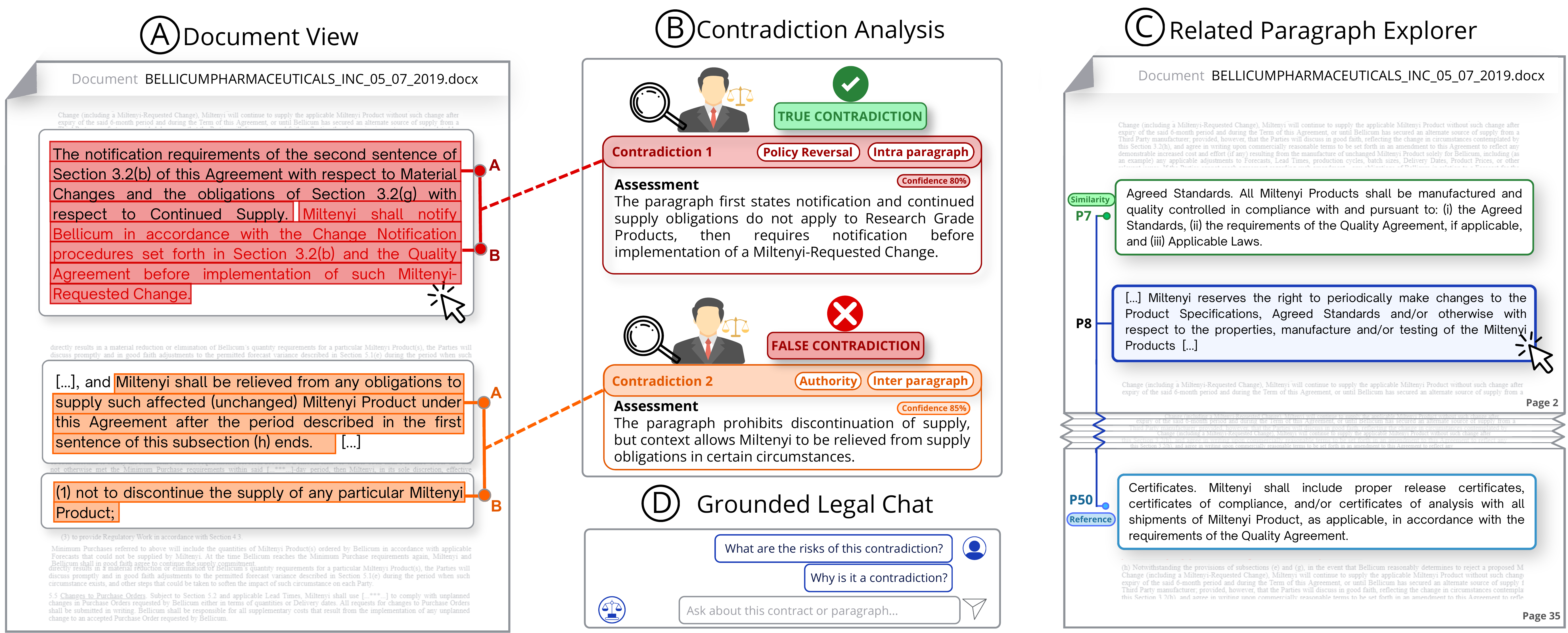}
    \caption{\systemname interface. (A) Document View highlights conflicting evidence spans in the contract text. (B) Contradiction Analysis lists candidates with category, scope, assessment, confidence, and evidence. (C) Related Paragraph Explorer brings distant but related clauses into a shared comparison space, distinguishing referential and semantic links. (D) Grounded Legal Chat supports explanation and follow-up reasoning grounded in the selected context.}
  \label{fig:ui-interface}
\end{figure*}
}

\newcommand{\systemreal}{
\begin{figure*}[t]
  \centering
  \includegraphics[width=0.95\linewidth]{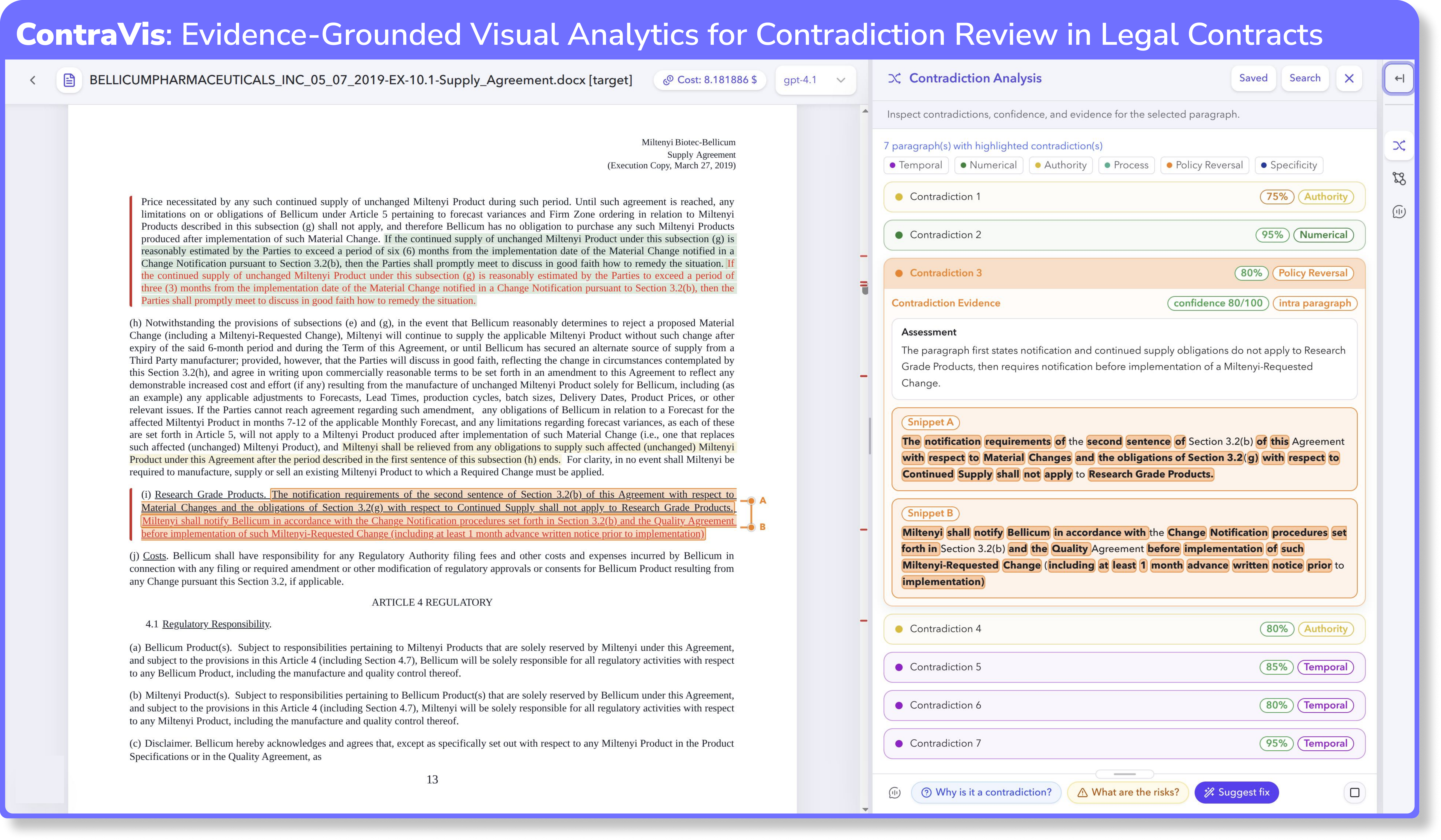}
  \caption{\systemname prototype interface on a CUAD contract: the \textit{Document View} (left) highlights contradiction evidence, and the \textit{Contradiction Analysis} panel (right) lists candidates with their category and confidence score, expanding one into its assessment and conflicting snippets (A, B).}
  \label{fig:systemreal}
\end{figure*}
}

\newcommand{\casestudytwo}{
\begin{figure}[t]
  \centering
  \includegraphics[width=\columnwidth]{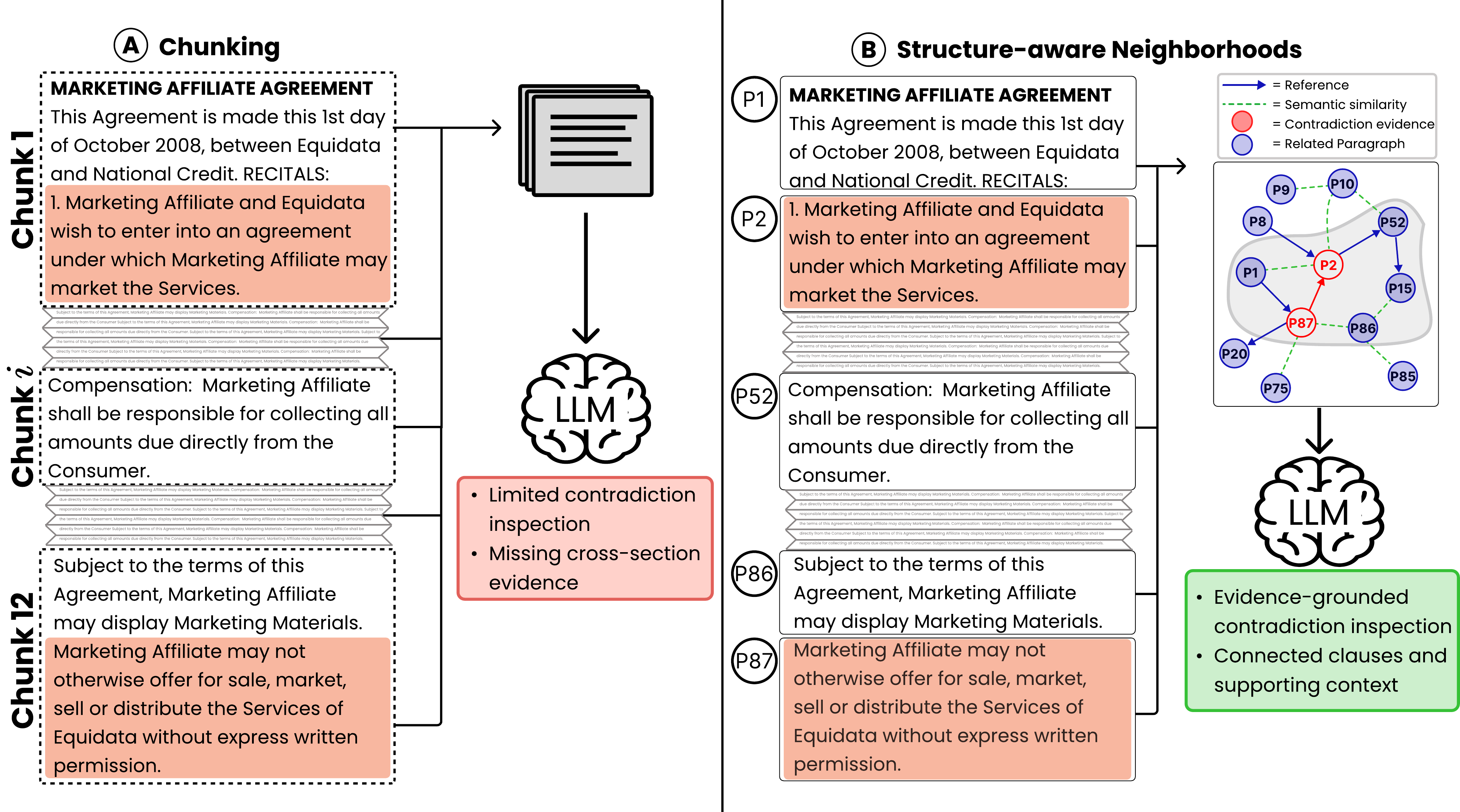}
  \vspace{-0.3cm}
  \caption{Fixed-size chunking vs. structure-aware neighborhoods. (A) Contradictory clauses far apart fall into different chunks, so cross-chunk contradictions are missed. (B) The typed paragraph graph reconnects them through referential and semantic links, enabling evidence-grounded analysis across sections.}
  \label{fig:casestudy2}
\end{figure}
}
\newcommand{\userevaluation}{
    \begin{table}[t]
    \caption{Usability and usefulness questionnaire results (five-point Likert scale).}
    \label{tab:likert-results}
    \centering
    \small
    \setlength{\tabcolsep}{10pt}
    \renewcommand{\arraystretch}{1.1}
    \rowcolors{1}{white}{rowgray}
        \begin{tabular}{@{}lccc@{}}
            \toprule
            \textbf{Question} & \textbf{P1} & \textbf{P2} & \textbf{P3} \\
            \midrule
            US1 & SA & SA & A \\
            US2 & SA & A  & A \\
            US3 & A  & A  & A \\
            UF1 & A  & A  & A \\
            UF2 & A  & SA & A \\
            UF3 & A  & N  & D \\
            \bottomrule
        \end{tabular}
        \smallskip
        \parbox{\linewidth}{\footnotesize SD: strongly disagree, D: disagree, N: neutral, A: agree, SA: strongly agree.}
    \end{table}
}

\newcommand{\casestudyone}{
    \rowcolors{3}{rowgray}{white}
    \begin{tabular}{@{}cccccccc@{}}
    \toprule
    \multirow{2}{*}{\textbf{Cont.}} &
    \multirow{2}{*}{\textbf{Pages}} &
    \multirow{2}{*}{\textbf{\(|V|\)}} &
    \multirow{2}{*}{\textbf{Inj.}} &
    \multicolumn{2}{c}{\textbf{LLM}} &
    \multicolumn{2}{c}{\textbf{\systemname}} \\
    \cmidrule(lr){5-6}\cmidrule(lr){7-8}
    & & & &
    \textbf{TP} & \textbf{FP} &
    \textbf{TP} & \textbf{FP} \\
    \midrule
    \(\mathcal{C}^{*}_{1}\) & 15 & 208 & 5 & 3 & 1 & 4 & 2 \\
    \(\mathcal{C}^{*}_{2}\) & 31 & 270 & 5 & 3 & 2 & 4 & 4 \\
    \(\mathcal{C}^{*}_{3}\) & 47 & 517 & 5 & 1 & 1 & 3 & 4 \\
    \(\mathcal{C}^{*}_{4}\) & 55 & 403 & 5 & 1 & 2 & 4 & 3 \\
    \bottomrule
\end{tabular}
}

\newcommand{\systemoverview}{
    \begin{table*}[t]
    \centering
    \caption{System requirements, analytical tasks, and interface support in \systemname.}
    \label{tab:design_goals}
    \footnotesize
    \setlength{\tabcolsep}{3pt}
    \renewcommand{\arraystretch}{1.05}
        \begin{tabularx}{\textwidth}{
            >{\raggedright\arraybackslash}p{0.34\textwidth}
            >{\raggedright\arraybackslash}X
            >{\raggedright\arraybackslash}p{0.22\textwidth}
        }
            \toprule
            \textbf{System Requirements} & \textbf{Analytical Tasks} & \textbf{Interface Support} \\
            \midrule
            
            \textbf{R1:} Structure-aware contradiction discovery in long contracts
            &
            \textbf{T1:} Triage contradiction-prone regions\newline
            \textbf{T2:} Inspect structure-aware neighborhood context\newline
            \textbf{T3:} Verify contradiction evidence and scope
            &
            Document View\newline
            Related Paragraph Explorer\newline
            Contradiction Analysis
            \\
            
            \midrule
            
            \textbf{R2:} Contextual inspection of potentially risky clauses
            &
            \textbf{T1:} Triage contradiction-prone regions\newline
            \textbf{T2:} Inspect structure-aware neighborhood context\newline
            \textbf{T4:} Explore grounded legal questions and interpretations
            &
            Document View\newline
            Related Paragraph Explorer\newline
            Grounded Legal Chat
            \\
            
            \midrule
            
            \textbf{R3:} Grounded legal interpretation and exploration
            &
            \textbf{T4:} Explore grounded legal questions and interpretations\newline
            \textbf{T5:} Interpret findings and explore corrective reasoning
            &
            Grounded Legal Chat\newline
            Contradiction Analysis
            \\
            
            \midrule
            
            \textbf{R4:} Evidence-grounded validation of contradiction findings
            &
            \textbf{T2:} Inspect structure-aware neighborhood context\newline
            \textbf{T3:} Verify contradiction evidence and scope\newline
            \textbf{T5:} Interpret findings and explore corrective reasoning
            &
            Contradiction Analysis\newline
            Related Paragraph Explorer
            \\
            
            \bottomrule
        \end{tabularx}
    \end{table*}
}

\newcommand{\tableRT}{
    \begin{table}[t]
        \centering
        \caption{System requirements and the analytical tasks.}
        \label{tab:req_tasks}
        \footnotesize
        \renewcommand{\arraystretch}{1.15}
        \setlength{\tabcolsep}{2pt}
        \rowcolors{2}{white}{rowgray}
        \begin{tabularx}{\columnwidth}{@{} X ccccc @{}}
            \toprule
            \textbf{System Requirements} & \textbf{T1} & \textbf{T2} & \textbf{T3} & \textbf{T4} & \textbf{T5} \\
            \midrule
            \textbf{R1:} Structure-aware contradiction discovery in contracts & \checkmark & \checkmark & \checkmark &            &            \\
            \textbf{R2:} Contextual inspection of potentially risky clauses        & \checkmark & \checkmark &            & \checkmark &            \\
            \textbf{R3:} Grounded legal interpretation and exploration             &            &            &            & \checkmark & \checkmark \\
            \textbf{R4:} Evidence-grounded validation of findings    &            & \checkmark & \checkmark &            & \checkmark \\
            \bottomrule
        \end{tabularx}
    \end{table}
}

\newcommand{\tableRI}{
    \begin{table}[t]
        \centering
        \caption{Interface components and analytical tasks in \systemname.}
        \label{tab:req_interface}
        \footnotesize
        \renewcommand{\arraystretch}{1.2}
        \setlength{\tabcolsep}{8pt}
        \rowcolors{2}{white}{rowgray}
        \begin{tabular}{@{}lccccc@{}}
            \toprule
            \textbf{Interface Component} & \textbf{T1} & \textbf{T2} & \textbf{T3} & \textbf{T4} & \textbf{T5} \\
            \midrule
            \textbf{Document View}              & \checkmark &            &            &            &            \\
            \textbf{Contradiction Analysis}     &            &            & \checkmark &            & \checkmark \\
            \textbf{Related Paragraph Explorer} &            & \checkmark &            &            &            \\
            \textbf{Grounded Legal Chat}        &            &            &            & \checkmark &            \\
            \bottomrule
        \end{tabular}
    \end{table}
}

\newcommand{\caseone}{
    \begin{table}[t]
        \caption{Tested documents and contradiction outcomes.}
        \label{tab:casestudyone}
        \centering
        \casestudyone
        \smallskip

        \footnotesize
        \(|V|\): number of paragraphs; Inj.: injected contradictions; TP: true positives; FP: false positives.
    \end{table}
}
\section{Introduction}

Legal contracts are high-stakes documents where small inconsistencies can create major financial and operational risks.
However, contract review remains largely manual and cognitively demanding, particularly for long agreements containing dense cross-references, repeated obligations, and domain-specific language~\cite{survey_legalcont_singh,cuad}.
Recent benchmarks further show that even strong LLMs still struggle with reliable legal reasoning in risk-sensitive contractual settings~\cite{contracteval}.
Prior work has advanced contract understanding through clause extraction, legal question answering, metadata identification, and document-level inference~\cite{nawar-etal-open,extraction_metadata,survey_legalcont_singh,contranli}.
At the same time, contradiction-oriented approaches have explored inconsistency detection in long documents~\cite{contradoc,chen-etal-2025-think,lattimer-etal-2023-fast} and logical relation classification between clause pairs~\cite{ichida_detecting_logical}.
However, these approaches operate as end-to-end predictors over isolated clause pairs or retrieved chunks: the context that conditions a prediction is not exposed to the analyst, so a flagged pair cannot be traced back to the contractual structure that makes it relevant.
In parallel, legal visualization systems demonstrate the value of structural exploration of legal corpora~\cite{legalvis,legalanalytics,viana}, but do not operationalize contradiction discovery in contracts~\cite{carvalho-transforming}.

We argue that contradiction analysis in legal contracts is a structure-aware analytical task rather than a clause-level classification problem.
To support this workflow, analysts must inspect related provisions, compare evidence across distant regions, and determine whether inconsistencies represent conflicts, exceptions, or refinements.

We present \systemname, a visual analytics system for human-in-the-loop contradiction analysis in legal contracts that combines graph-conditioned LLM reasoning with multiple views.
The system represents contracts as typed paragraph graphs combining explicit references with semantic relationships between paragraphs.
\systemname uses this graph as both a computational substrate for contradiction analysis and a visual structure for contextual inspection.
Following LegalWiz~\cite{legalwiz}, contradictions are typed as \textit{Temporal}, \textit{Numerical}, \textit{Authority}, \textit{Process}, \textit{Policy Reversal}, or \textit{Specificity}, and scoped as intra- or inter-paragraph (\appref{app:taxonomy}).
\systemname supports contract triage, paragraph-neighborhood exploration, verification, and evidence-grounded legal interpretation.

We frame this work as an early-stage design study~\cite{sedlmair_designstudy}: its contributions lie at the problem-characterization and task-abstraction levels of visualization design rather than in new visual encodings.
This work makes three contributions.
First, we formulate contradiction review as an evidence-grounded, human-in-the-loop visual analytics task, grounded in lawyer-centered requirements and abstracted into analytical tasks, rather than automatic clause-pair classification.
Second, we propose a typed paragraph graph representation with a dual role: the structure that conditions LLM reasoning is also the interactive representation the analyst explores, keeping model context and human inspection aligned---unlike standard RAG pipelines, where the conditioning context stays hidden from the analyst.
Third, we present \systemname, an interactive system integrating document reading, evidence comparison, paragraph-neighborhood exploration, and grounded legal assistance, together with a formative expert evaluation and a controlled comparison showing that graph-conditioned reasoning recovers more injected contradictions as contracts grow, while surfacing additional candidates for analyst validation and design implications for evidence-grounded, LLM-assisted document review.
Code, prompts, and evaluation material are available at \url{https://visualdslab.com/papers/ContraVis}.

\tableRT
\section{Related Work}

\subsection{Legal Contract Understanding with NLP and LLMs}

Research on legal contract analysis evolved from clause extraction benchmarks and expert-annotated datasets toward broader reasoning and understanding tasks~\cite{cuad,survey_legalcont_singh}.
ContractNLI showed that document-level inference over contracts remains challenging for language models, particularly in long-document settings~\cite{contranli}.
Subsequent work explored metadata extraction, legal question answering, summarization, and contract understanding applications~\cite{nawar-etal-open,extraction_metadata,sancheti_what_read}.
Recent studies show that structural and formatting cues strongly influence legal LLM behavior and that preserving document structure aids evidence localization and legal reasoning~\cite{braun_hiddenstructure,ruiyu,LegalBench}.
Although these approaches improve legal language understanding, they operate as end-to-end predictors, offering analysts no structural context for inspecting evidence or validating contradictions.

\subsection{Contradiction Detection and Retrieval-Augmented Analysis}
Contradiction analysis has been explored across factual, regulatory, and legal domains~\cite{detection_schumann}.
In legal contracts, prior work formulates it as logical relation classification between clause pairs using natural language inference~\cite{ichida_detecting_logical}.
Recent studies investigated self-contradiction detection~\cite{contradoc}, reference-aware contradiction discovery~\cite{chen-etal-2025-think}, chunk-based inconsistency analysis~\cite{lattimer-etal-2023-fast}, and multi-agent contradiction workflows~\cite{legalwiz}.
In parallel, graph-aware retrieval and reranking show that modeling document connectivity improves evidence discovery and contextual reasoning~\cite{re2g,jialin_dontforgetconnectimproving,mesgar_graph_coherence}.
\systemname draws on both lines but differs in what the analyst sees: whereas these approaches predict over isolated clause pairs or chunks with opaque retrieved context, our candidates remain traceable to verbatim evidence spans and inspectable graph neighborhoods, supporting expert validation rather than only detection.

\subsection{Visual and Interactive Analysis of Legal Documents}
Prior visualization and visual analytics systems show that structural relationships, citation networks, and argument connections matter for legal reasoning workflows~\cite{legalvis,viana,carvalho-transforming}.
More recent research in legal HCI advocates moving beyond static document-centric workflows toward systems that support exploration, interpretation, and interactive reasoning over legal content~\cite{living_contracts}.
Visual analytics systems have also begun coupling graph structure with LLM reasoning to keep model-generated explanations inspectable~\cite{graphllm_anomalies}.
Whereas these systems support structural exploration of legal corpora, \systemname couples the visual structure to the reasoning process: the typed paragraph graph shown to the analyst is the same structure that conditions LLM inference, aligning model context with human inspection.

\section{System Overview}
\label{sec:systemoverview}

\subsection{System Requirements and Analytical Tasks}
Requirements were elicited through semi-structured discussions with $2$ contract-domain lawyers, who described their review practice and the difficulty of tracing obligations across distant provisions.
From these we distilled four lawyer-centered requirements (R1--R4), abstracted into five analytical tasks (T1--T5) mapped in \autoref{tab:req_tasks}:
\textbf{T1}: triage contradiction-prone regions;
\textbf{T2}: inspect structure-aware neighborhood context;
\textbf{T3}: verify contradiction evidence and scope;
\textbf{T4}: explore grounded legal questions and interpretations; and
\textbf{T5}: interpret findings and explore corrective reasoning.

\subsection{Workflow}
\autoref{fig:workflow} summarizes the human-in-the-loop workflow: from a legal contract (A), \systemname constructs a typed paragraph graph (B), retrieves local neighborhoods as contextual evidence for graph-conditioned reasoning (C), and surfaces contradiction candidates with supporting evidence, which analysts inspect through four coordinated views (D, \autoref{sec:system}).

\section{Graph-Conditioned Contradiction Analysis}
\label{sec:methodology}

This section describes the graph representation, controlled contradiction-generation process, and graph-conditioned reasoning strategy used in \systemname.

\figworkflow

\subsection{Graph Representation}
\label{sec:base-graph-construction}

Given a contract \(\mathcal{C}=\{p_1,\dots,p_n\}\), we construct a typed paragraph graph
$
\mathcal{G}=(V,E_r,E_s),
$
where each node \(v_i \in V\) represents a paragraph \(p_i\).

\myparagraph{\textbf{Referential edges} (\(E_r\)).}
Edges connecting paragraphs through explicit contractual cross-references, identified with rule-based patterns over legal identifiers (\eg, ``Section 3.2'').

\myparagraph{\textbf{Semantic edges} (\(E_s\)).}
Edges connecting semantically related paragraphs whose \texttt{all-MiniLM-L6-v2} embeddings exceed a cosine similarity of \(\tau = 0.80\).

Referential links preserve the document's explicit legal structure, whereas semantic links capture latent topical relationships. Construction rules, preprocessing, and parameter settings are described in \appref{app:graph-construction}.

\uiinterface

\subsection{Controlled Evaluation Dataset}

To study contradiction inspection under controlled conditions, we construct synthetic contradiction-analysis scenarios from contracts in the CUAD corpus~\cite{cuad}.
Starting from the typed paragraph graph \(\mathcal{G}\), structurally related paragraph neighborhoods are provided as contextual input to \texttt{GPT-4.1}, which generates controlled inconsistencies using a taxonomy-conditioned prompt derived from LegalWiz~\cite{legalwiz} (\appref{app:prompts}).
The generated contradictions are inserted into compatible locations of the original contract while preserving the remaining document structure, producing a synthetic contradicted contract \(\mathcal{C}^{*}\).
Each candidate is kept only when its NLI contradiction score satisfies \(\gamma \geq 0.70\) and its per-word perplexity increase satisfies \(\Delta\mathrm{PPL}_{\text{norm}} < \delta = 0.01\) (\appref{app:synthetic-generation}); accepted candidates are then manually reviewed by the researchers for contextual plausibility, consistency with the intended category, and alignment with the contractual regions used during generation.
Contradiction locations are known to the experimenters but hidden from both the system and analysts during inspection.
For each contract, five contradictions are targeted. Candidates are filtered with NLI and perplexity, manually reviewed, and rejected ones are regenerated until five validated contradictions are obtained.

\subsection{Structure-Aware Analytical Reasoning}

During analysis, the system reconstructs the typed paragraph graph \(\mathcal{G}^{*}\) from the input contract and retrieves local graph neighborhoods for contextual inspection (\appref{app:detection}).
Neighborhoods are limited to direct (1-hop) graph connections, combining all referentially linked paragraphs and the top-5 semantically similar neighbors associated with the selected paragraph.
Contextualizing each paragraph through both referential and semantic relationships lets candidates be surfaced from structurally connected regions together with evidence spans, contradiction categories, and contextual relationships (\appref{app:prompts}).
Rather than ranking candidates as final decisions, \systemname presents them as analytical evidence for expert inspection, validation, and interpretation.

\section{\systemname System}
\label{sec:system}

\autoref{fig:ui-interface} presents the four coordinated views that support
contradiction-centered contract review in \systemname; each view is annotated
below with the analytical tasks it supports.

\subsection{Document View}
\label{sec:documentview}

The \textit{Document View} (\autoref{fig:ui-interface}.A) supports close reading and triage (T1) in long contracts.
Because the contract text is the legal source of truth, the document---rather than a derived abstraction---is the primary representation in which analysts read and judge provisions.
The view presents the contract as an ordered sequence of paragraphs linked to the typed paragraph graph; contradiction candidates and related paragraphs are highlighted in the text via linked highlighting, and selecting a paragraph updates the graph neighborhood, contradiction evidence, and assistant context across the other views.

\subsection{Contradiction Analysis}
\label{sec:contradictionanalysis}

The \textit{Contradiction Analysis} panel (\autoref{fig:ui-interface}.B) supports evidence inspection and candidate validation (T3, T5), exposing candidate inconsistencies together with supporting evidence rather than final predictions---reflecting the legal nature of the task, in which potential conflicts must be interpreted in context before they count as meaningful inconsistencies.
For each candidate, the panel presents:
(1) the contradiction category,
(2) a brief assessment,
(3) contradiction scope (intra- or inter-paragraph), 
(4) the model-reported confidence score (0--100), and
(5) highlighted evidence spans associated with the candidate.
Evidence snippets are synchronized with the \textit{Document View}, letting analysts cross-check conflicting provisions in their original contractual context.
Users can mark each candidate as a true contradiction, a false contradiction, or a contextual exception/refinement, and the panel connects to the \textit{Grounded Legal Chat} to request explanations, inspect potential risks, and explore revision strategies for the selected candidate.

\subsection{Related Paragraph Explorer}
\label{sec:relatedparagraph}

The \textit{Related Paragraph Explorer} (\autoref{fig:ui-interface}.C) supports neighborhood exploration (T2) by externalizing the graph neighborhood of a selected paragraph \(v_i\), distinguishing referential edges \(E_r\) from semantic edges \(E_s\).
We adopt a node-link representation because neighborhoods are small in practice (1-hop referential links plus top-5 semantic neighbors): at this scale the diagrams remain legible, preserve per-paragraph coordination with the \textit{Document View}, and map edge types to link encodings, which matrices would obscure.
A key interaction of this view is proximity-based comparison,
motivated by the pairwise nature of contradiction validation: conflicting provisions must be read together yet may lie far apart.
When analysts select a paragraph or contradiction candidate, \systemname brings distant but structurally related paragraphs into a shared comparison space, enabling side-by-side inspection of evidence without navigating across contract sections.

\subsection{Grounded Legal Chat}
\label{sec:legalchat}
The \textit{Grounded Legal Chat} (\autoref{fig:ui-interface}.D) supports grounded legal interpretation and follow-up reasoning (T4).
Unlike a generic conversational assistant, the chat is conditioned on the selected paragraph, its graph neighborhood, and the current contradiction evidence---the same context that conditions detection---so conversational answers stay anchored to inspectable evidence rather than free-form model output.
Analysts can ask global or targeted questions about clauses and contradiction candidates, supporting explanation, risk inspection, and revision-oriented interactions.
\section{Controlled Comparison and Usage Scenario}

\label{sec:case-studies}

\subsection{Controlled Comparison with Standalone LLM Analysis}

We compare \systemname against standalone LLM analysis---the same detection model and prompt structure operating on the contract text without graph conditioning---on four contracts of increasing length (15--55 pages; 208--517 paragraphs), each containing five injected contradictions (\(\mathcal{C}^{*}\)).
The comparison is deliberately small in scale: rather than a benchmark, it functions as an ablation that isolates the contribution of the structural component precisely where it matters most, long contracts.
\autoref{tab:casestudyone} reports per-document true and false positives; matching is automatic at the evidence-span level, with a surfaced candidate counting as a true positive when its evidence spans overlap an injected contradiction, and as a false positive otherwise.

\caseone

On the shorter contracts, both approaches recover most injected contradictions (3--4 of 5).
As length grows, however, standalone recovery degrades to 1 of 5, while graph-conditioned reasoning still recovers 3--4 of 5.
The failure mode is consistent: when a contradiction depends on distant or semantically related provisions, the standalone model produced weakly contextualized candidates, whereas graph neighborhoods keep the relevant regions jointly visible to the model and, through the coordinated views, to the analyst---who can compare evidence spans, inspect related paragraphs, and understand why a contractual region is relevant to a surfaced candidate.

Graph conditioning also surfaces more candidates beyond the injected ones (2--4 versus 1--2 per contract), reflecting the ambiguity of legal language and the deliberately high-recall detection bias (\appref{app:prompts}).
In legal review this trade-off is asymmetric: a missed contradiction is a silent legal risk, whereas an additional candidate costs bounded review effort.
Consistent with the system's human-in-the-loop design, \systemname does not treat these cases as final detection errors but exposes them as inspectable candidates (each with evidence spans and a confidence score) that analysts can validate, contextualize, or discard through coordinated visual inspection (\autoref{fig:ui-interface}.B).

\subsection{Chunking Limitations and Structure-Aware Neighborhoods}

Large contracts often contain dependencies distributed across distant sections of the document. 
Although fixed-size chunking reduces context-window overload during LLM processing, it fragments structural and semantic relationships that analysts must inspect when evaluating supporting evidence and potential contradictions.
By preserving these relationships in the typed paragraph graph, \systemname supports graph-conditioned reasoning and interactive exploration, allowing analysts to compare evidence across distant contractual regions and inspect the context underlying contradiction candidates.
\autoref{fig:casestudy2} illustrates this contrast: fixed-size chunking may separate related clauses into independent contexts, whereas structure-aware neighborhoods reconnect them through referential and semantic relationships.

\casestudytwo
\section{Formative Expert Evaluation}
\label{sec:evaluation}
We conducted a formative expert evaluation with three contract-domain experts to assess whether \systemname supports meaningful contradiction inspection, evidence interpretation, and structure-aware exploration in legal-review workflows. 
Rather than aiming for statistically generalizable usability measures, the evaluation focused on collecting domain-informed feedback about the usefulness, interpretability, and limitations of the proposed visual analytics workflow. 
This evaluation is appropriate for an early-stage design study because access to contract-specialized legal reviewers is limited and their qualitative feedback is critical for refining the design.

\subsection{Participants}
\label{sec:participants}
The study involved three contract-domain lawyers from a large Brazilian law firm.
None of them had participated in the requirements-elicitation discussions (\autoref{sec:systemoverview}), so the evaluation was not performed by the experts who informed the design.
All participants had direct experience with contract drafting and review.
P1 had 6--10 years of contract-related experience and worked with contracts several times per week. 
P2 and P3 had 3--5 years of experience; P2 worked with contracts several times per month, while P3 worked with contracts several times per week.

\subsection{Procedure}
\label{sec:procedure}

The evaluation was conducted through live in-person sessions, each lasting approximately 30 minutes.
Each session began with an introduction to \systemname, including the typed paragraph graph representation, contradiction-analysis workflow, and coordinated visual interface. 
Participants then interacted with the system through two study tasks (S1--S2, distinct from the analytical tasks T1--T5), thinking aloud about why each surfaced candidate was a correct or incorrect contradiction. Sessions were not recorded, but the experimenters noted participants' feedback.

\myparagraph{\textbf{S1}: Guided contradiction inspection.}
Participants explored a pre-selected \(\mathcal{C}^{*}\) contract containing an injected contradiction involving late fees and bounced-check penalties.
They inspected supporting evidence, compared contractual fragments, and explored relationships among the involved clauses.

\myparagraph{\textbf{S2}: Free exploration of long contracts.}
Participants selected contracts ranging from 15 to 55 pages and freely explored contradiction candidates surfaced by the system, focusing on cases they considered legally relevant.

Following the exploration tasks, participants answered usability, usefulness, and contradiction-analysis questions using a five-point Likert scale (\appref{app:user-questionnaire}) and provided open-ended feedback about useful features, workflow limitations, and possible improvements.

\subsection{Questionnaire Results}
\label{sec:questionnaire}
Responses to the usability items were uniformly positive (agree or strongly agree on US1--US3), as were the usefulness of the tool for supporting real contract review (UF1) and the value of the contradiction explanations for legal validation (UF2).
Adoption intent, however, diverged: asked whether they would use \systemname in their own workflow (UF3), P1 agreed, P2 was neutral, and P3 disagreed, preferring to stay with an AI legal-assistance tool already integrated into their firm's workflow (\appref{app:user-questionnaire}).
This spread is consistent with the experts' overall assessment: they found the workflow useful for contradiction inspection, evidence interpretation, and structure-aware exploration, while noting that additional refinement is necessary before deployment in professional legal settings.

\subsection{Qualitative Findings}
\label{sec:findings}

\myparagraph{Finding 1: Visual evidence comparison supported contradiction validation.}
Participants consistently identified contradiction-centered visual exploration as one of the most useful aspects of \systemname.
The ability to compare conflicting spans directly in context helped participants inspect and validate contradiction candidates more effectively than isolated textual outputs.

\myparagraph{Finding 2: Structure-aware exploration supported contextual inspection, but legal relationship types require refinement.}
Participants found value in exploring relationships between distant contractual provisions.
However, one participant noted that the system occasionally classified rule-exception relationships as contradictions.
This observation is consistent with the additional candidates surfaced in the controlled comparison (\autoref{sec:case-studies}) and suggests a concrete source for them:
future versions should better distinguish genuine contradictions from legal exceptions, refinements, and scope restrictions.

\myparagraph{Finding 3: Grounded explanations and revision support were useful, but require deeper legal integration.}
Participants valued the explanatory and corrective capabilities of the system, especially for understanding why candidates were flagged and how contractual language could be revised.
However, these features should remain tightly grounded in evidence and legal context to avoid over-reliance on LLM-generated reasoning.

\subsection{Design Implications}
\label{sec:designimplications}

The evaluation suggests several implications for improving \systemname.
First, visual evidence comparison should remain central to the workflow.
Second, the relationship model should better distinguish contradictions from legal exceptions and scope refinements.
Third, explanation and rewriting features should be strengthened while preserving explicit evidence grounding and expert control over interpretation.
Fourth, the divergent adoption responses suggest that usefulness for inspection does not by itself translate into workflow adoption; integration with existing review practices deserves design attention of its own.

\section{Discussion, Limitations, and Future Work}
\myparagraph{Discussion.}
Our observations suggest that contradiction analysis in long contracts should not be treated as a purely long-context prompting problem.
Although modern LLMs reason effectively over localized regions, their behavior may degrade when contradictions depend on relationships distributed across distant sections of large agreements.
Rather than uniformly increasing context size, preserving structure-aware neighborhoods appears more effective for evidence-grounded inspection and human validation.
Our expert study suggests that coordinated evidence views, rather than model accuracy alone, drive trustworthy review: in-context comparison was how experts accepted, contextualized, or discarded candidates.
The per-candidate confidence score (0--100) (\autoref{sec:contradictionanalysis}) is self-reported by the model and not calibrated, so analysts should read it as a triage cue rather than a probability; calibrating it against expert validation decisions, and encoding graded signals such as edge similarities and reranking scores (\appref{app:prompts}), remain directions for improving trust calibration.

\myparagraph{Limitations.}
Our formative study involved three experts from a single Brazilian law firm, so the results do not support generalizable conclusions about usability or effectiveness.
Our evaluation also relies on synthetic contradictions---no public benchmark of naturally occurring document-level contractual contradictions exists, and ContractNLI~\cite{contranli} targets clause-pair entailment---which, despite graph-conditioned generation, automatic validation, and manual review (\appref{app:synthetic-generation}), may not fully capture the ambiguity of real disputes.
The comparison covers four contracts and one baseline, isolating the effect of graph conditioning without characterizing it across corpora or models.
Graph-conditioned reasoning also remains sensitive to paragraph segmentation, embedding quality, similarity thresholds, and prompt design.
Finally, as an early-stage design study, \systemname has not been deployed in professional practice; the deployment and longitudinal-reflection stages of design-study methodology~\cite{sedlmair_designstudy} remain open.

\myparagraph{Future Work.}
Future work includes enriching the typed paragraph graph with contract-level entities and obligations, expanding toward broader contract-quality tasks such as ambiguity and redundancy inspection, and improving robustness against adversarial content.
We also plan sensitivity analyses of key graph parameters (\eg, threshold \(\tau\)), component ablations of referential edges, semantic edges, and reranking, experiments on larger collections, and longitudinal studies with legal professionals assessing trust calibration, usability, task-based analyst efficiency and cognitive workload against text-only baselines, and naturally occurring inconsistencies.
\section{Conclusion}
We presented \systemname, a visual analytics system for structure-aware contradiction analysis in legal contracts.
The system models contracts as typed paragraph graphs whose dual role---conditioning LLM reasoning while serving as the interactive representation the analyst explores---keeps model context and human inspection aligned.
In a controlled comparison, graph-conditioned reasoning recovered more injected contradictions than standalone LLM analysis as contract length grew, and our formative study with contract-domain lawyers indicated that in-context evidence comparison, rather than model output alone, is what supported expert validation.
Overall, \systemname frames contradiction analysis as a human-in-the-loop workflow that combines automated reasoning, evidence-grounded visual exploration, and expert legal judgment ---a design we expect to transfer to other long, densely interlinked documents such as regulations and policy corpora.

\bibliographystyle{IEEEtran}

\bibliography{refs}

@inproceedings{contranli,
    title = "{C}ontract{NLI}: A Dataset for Document-level Natural Language Inference for Contracts",
    author = "Koreeda, Yuta  and Manning, Christopher",
    booktitle = "Findings of EMNLP",
    year = "2021",
    doi = "10.18653/v1/2021.findings-emnlp.164",
}

@article{legalanalytics,
  author  = {Resck, Lucas and Moreno-Vera, Felipe A. and Veiga, Tobias and Paucar, Gerardo and
             Fajreldines, Ezequiel and Klafke, Guilherme and Nonato, Luis Gustavo and Poco, Jorge},
  title   = {LegalAnalytics: Bridging Visual Explanations and Workload Streamline in Brazilian Supreme Court Appeals},
  journal = {Artificial Intelligence and Law},
  year    = {2025}
}

@article{sedlmair_designstudy,
  author  = {Sedlmair, Michael and Meyer, Miriah and Munzner, Tamara},
  title   = {Design Study Methodology: Reflections from the Trenches and the Stacks},
  journal = {IEEE TVCG},
  volume  = {18}, number = {12}, pages = {2431--2440}, year = {2012}
}

@misc{cuad,
    title={CUAD: An Expert-Annotated NLP Dataset for Legal Contract Review}, 
    author={Dan Hendrycks and Collin Burns and Anya Chen and Spencer Ball},
    year={2021},
    eprint={2103.06268},
}

@inproceedings{contracteval,
    title = "{C}ontract{E}val: Benchmarking {LLM}s for Clause-Level Legal Risk Identification in Commercial Contracts",
    author = "Liu, Shuang  and Li, Zelong  and Ma, Ruoyun  and Zhao, Haiyan  and Du, Mengnan",
    booktitle = "NLLP Workshop",
    year = "2025",
    doi = "10.18653/v1/2025.nllp-1.19",
    pages = "291--291",
 }

@article{survey_legalcont_singh,
    title={A survey of classification tasks and approaches for legal contracts},
    author={Singh, Amrita and Joshi, Aditya and Jiang, Jiaojiao and Paik, Hye-young},
    journal={Artificial Intelligence Review},
    volume={58},
    number={12},
    pages={380},
    year={2025},
    doi={10.1007/s10462-025-11359-8}
}

@inproceedings{nawar-etal-open,
    title = "An Open Source Contractual Language Understanding Application Using Machine Learning",
    author = "Nawar, Afra  and Rakib, Mohammed  and Hai, Salma Abdul  and Haq, Sanaulla",
    booktitle = "LT4FISS @ LREC",
    year = "2022",
    pages = "42--50",
}

@misc{extraction_metadata,
    title={Metadata Extraction Leveraging Large Language Models}, 
    author={Cuize Han and Sesh Jalagam},
    year={2025},
    eprint={2510.19334}, 
}

@misc{braun_hiddenstructure,
    title={The Hidden Structure -- Improving Legal Document Understanding Through Explicit Text Formatting}, 
    author={Christian Braun and Alexander Lilienbeck and Daniel Mentjukov},
    year={2025},
    eprint={2505.12837},
}

@inproceedings{sancheti_what_read,
    title = "What to Read in a Contract? Party-Specific Summarization of Legal Obligations, Entitlements, and Prohibitions",
    author = "Sancheti, Abhilasha and Garimella, Aparna  and Srinivasan, Balaji  and Rudinger, Rachel",
    booktitle = "EMNLP",
    year = "2023",
    doi = "10.18653/v1/2023.emnlp-main.909",
}

@INPROCEEDINGS{detection_schumann,
    author={Schumann, Gerrit and Gómex, Jorge Marx},
    booktitle = "IDSTA",
    title={Detection of Conflicts, Contradictions and Inconsistencies in Regulatory Documents: A Literature Review}, 
    year={2024},
    pages={81-88},
    doi={10.1109/IDSTA62194.2024.10747003}
}

@misc{ichida_detecting_logical,
    title={Detecting Logical Relation In Contract Clauses}, 
    author={Alexandre Yukio Ichida and Felipe Meneguzzi},
    year={2021},
    eprint={2111.01856},
}

@inproceedings{lattimer-etal-2023-fast,
    title = "Fast and Accurate Factual Inconsistency Detection Over Long Documents",
    author = "Lattimer, Barrett  and Chen, Patrick H.  and Zhang, Xinyuan  and Yang, Yi",
    booktitle = "EMNLP",
    year = "2023",
    doi = "10.18653/v1/2023.emnlp-main.105",
}

@inproceedings{chen-etal-2025-think,
    title = "Think Wider, Detect Sharper: Reinforced Reference Coverage for Document-Level Self-Contradiction Detection",
    author = "Chen, Yuhao  and Lyu, Yuanjie  and Liu, Shuochen  and Zhang, Chao  and Lv, Junhui  and Xu, Tong",
    booktitle = "Proc. EMNLP",
    year = "2025",
    doi = "10.18653/v1/2025.emnlp-main.67",
    pages = "1273--1288",
}

@inproceedings{carvalho-transforming,
    author = {Carvalho, Nuno Ramos and Barbosa, Lu\'{\i}s Soares},
    title = {Transforming Legal Documents for Visualization and Analysis},
    year = {2018},
    doi = {10.1145/3209415.3209424},
    booktitle = "ICEGOV 2018"
}

@inproceedings{contradoc,
    title = "{C}ontra{D}oc: Understanding Self-Contradictions in Documents with Large Language Models",
    author = "Li, Jierui and Raheja, Vipul  and Kumar, Dhruv",
    booktitle = "NAACL-HLT",
    year = "2024",
    doi = "10.18653/v1/2024.naacl-long.362",
    pages = "6509--6523",
}

@misc{legalwiz,
    title={LegalWiz: A Multi-Agent Generation Framework for Contradiction Detection in Legal Documents}, 
    author={Ananya Mantravadi and Shivali Dalmia and Olga Pospelova and Abhishek Mukherji and Nand Dave and Anudha Mittal},
    year={2025},
}

@inproceedings{graphllm_anomalies,
  author    = {Heredia, Juanpablo and Estrada-Rayme, Leighton and Matos-Cangalaya, Jeremy and Poco, Jorge},
  title     = {Interactive Exploration and Explanation of Spatiotemporal Anomalies with Graph-LLM Integration},
  booktitle = {SIBGRAPI},
  year      = {2025}
}

@ARTICLE{legalvis,
    author={Resck, Lucas E. and Ponciano, Jean R. and Nonato, Luis Gustavo and Poco, Jorge},
    journal={IEEE TVCG},
    title={LegalVis: Exploring and Inferring Precedent Citations in Legal Documents}, 
    year={2023},
    volume={29},
    number={6},
    pages={3105-3120},
    doi={10.1109/TVCG.2022.3152450}
}

@INPROCEEDINGS{viana,
    author={Sperrle, Fabian and Sevastjanova, Rita and Kehlbeck, Rebecca and El-Assady, Mennatallah},
    booktitle = "VAST",
    title={VIANA: Visual Interactive Annotation of Argumentation}, 
    year={2019},
    doi={10.1109/VAST47406.2019.8986917}
}

@inproceedings{re2g,
    title = "{R}e2{G}: Retrieve, Rerank, Generate",
    author = "Glass, Michael  and Rossiello, Gaetano  and Chowdhury, Md Faisal Mahbub  and Naik, Ankita  and Cai, Pengshan  and Gliozzo, Alfio",
    booktitle = "NAACL-HLT",
    year = "2022",
    doi = "10.18653/v1/2022.naacl-main.194",
}

@misc{jialin_dontforgetconnectimproving,
    title={Don't Forget to Connect! Improving RAG with Graph-based Reranking}, 
    author={Jialin Dong and Bahare Fatemi and Bryan Perozzi and Lin F. Yang and Anton Tsitsulin},
    year={2024},
    eprint={2405.18414},
}

@inproceedings{mesgar_graph_coherence,
    title = "A Neural Graph-based Local Coherence Model",
    author = "Mesgar, Mohsen and Ribeiro, Leonardo F. R.  and Gurevych, Iryna",
    booktitle = "Findings of EMNLP",
    year = "2021",
    doi = "10.18653/v1/2021.findings-emnlp.199",
}

@inproceedings{living_contracts,
    author = {Huang, Ziheng and Kar, Robin and Sundaram, Hari and August, Tal},
    title = {Living Contracts: Beyond Document-Centric Interaction with Legal Agreements},
    year = {2026},
    doi = {10.1145/3772318.3791386},
    booktitle = "CHI",
}

@misc{LegalBench,
      title={LegalBench-RAG: A Benchmark for Retrieval-Augmented Generation in the Legal Domain}, 
      author={Nicholas Pipitone and Ghita Houir Alami},
      year={2024},
}

@misc{ruiyu,
      title={SF-RAG: Structure-Fidelity Retrieval-Augmented Generation for Academic Question Answering}, 
      author={Rui Yu and Tianyi Wang and Ruixia Liu and Yinglong Wang},
      year={2026}
}

\clearpage

\appendix

\section{Appendix}

\systemreal

\subsection{Contradiction Taxonomy}
\label{app:taxonomy}

Following the contradiction taxonomy proposed in LegalWiz~\cite{legalwiz}, the generated contradictions cover six categories:
(i) \textbf{Temporal}, involving dates, deadlines, durations, or event ordering;
(ii) \textbf{Numerical}, concerning monetary values, percentages, or quantities;
(iii) \textbf{Authority}, involving the responsible entity or authoritative source;
(iv) \textbf{Process}, concerning operational procedures or execution conditions;
(v) \textbf{Policy Reversal}, representing direct negation of previously established rules or obligations; and
(vi) \textbf{Specificity}, involving differences in scope or granularity.
\subsection{Graph Construction Details}
\label{app:graph-construction}

\myparagraph{Preprocessing.}
Non-contractual and low-information paragraphs (section headers, repetitive footnotes) are removed using structural heuristics.
Each remaining paragraph becomes a node, \(V=\{v_1,\dots,v_n\}\).

\myparagraph{Referential edges.}
A directed edge \((v_i, v_j) \in E_r\) is added when \(p_i\) explicitly references \(p_j\).
Extraction proceeds in three steps:
(1) regular expressions capture legal identifiers over section, article, clause, and subclause numbering schemes (\eg, ``Section 3.2'', ``Article 5'', ``Section 4.1(a)'');
(2) each identifier is resolved to its target paragraph by matching against the numbering recovered during preprocessing; and
(3) contextual filtering discards self-references and structural heading matches.

\myparagraph{Semantic edges.}
Each paragraph is embedded as \(e_i = \phi(p_i)\), where \(\phi(\cdot)\) is the \texttt{all-MiniLM-L6-v2} SentenceTransformer model.
An edge \((v_i, v_j, s_{ij}) \in E_s\) is added when \(s_{ij} = \cos(e_i, e_j) \geq \tau\), with \(\tau = 0.80\).

The resulting hybrid graph combines explicit legal structure (\(E_r\)) with latent topical structure (\(E_s\)).
\subsection{Synthetic Contradiction Generation and Validation}
\label{app:synthetic-generation}

Starting from the preprocessing graph \(\mathcal{G}=(V,E_r,E_s)\), we generate controlled contradictions through the four-stage pipeline summarized in \autoref{alg:synthetic}.

\begin{algorithm}[h]
\begin{algorithmic}[1]
\REQUIRE preprocessing graph \(\mathcal{G}=(V,E_r,E_s)\)
\ENSURE synthetic contradicted contract \(\mathcal{C}^{*}\)
\STATE \(H \gets \operatorname*{Top}_{m}\{\, d(v) : v \in V \,\}\) \COMMENT{seed selection, \(m=5\)}
\FORALL{seed \(v \in H\)}
    \FORALL{neighbor \(n \in N(v)\)}
        \STATE \(\mathrm{score}[n] \gets \operatorname{CrossEncoder}(v, n)\) \COMMENT{pairwise reranking}
    \ENDFOR
    \STATE \(N_K(v) \gets \operatorname{TopK}(\mathrm{score}, K)\) \COMMENT{\(K=5\)}
    \STATE \((s, c) \gets \operatorname{LLM}(v, N_K(v))\) \COMMENT{taxonomy-conditioned prompt}
    \STATE \(\gamma \gets P_{\mathrm{NLI}}(\text{contradiction} \mid s, c)\)
    \STATE \(\Delta\mathrm{PPL}_{\text{norm}} \gets (\mathrm{PPL}(p^{*}) - \mathrm{PPL}(p)) / |c|\) \COMMENT{per-word normalized}
    \IF{\(\gamma \geq \alpha\) \AND \(\Delta\mathrm{PPL}_{\text{norm}} < \delta\)}
        \STATE insert \(c\) into its target paragraph in \(\mathcal{C}^{*}\) \COMMENT{\(\alpha=0.70,\ \delta=0.01\)}
    \ENDIF
\ENDFOR
\STATE expert manual review of accepted contradictions
\end{algorithmic}
\caption{Synthetic contradiction generation and validation.}
\label{alg:synthetic}
\end{algorithm}

\myparagraph{Stages.}
(1)~\emph{Seed selection} picks the \(m\) highest-degree nodes of the hybrid graph.
(2)~\emph{Neighborhood reranking} scores each candidate neighbor with the cross-encoder \texttt{ms-marco-MiniLM-L-6-v2} and keeps the top \(K\) reference and semantic neighbors, duplicates removed.
(3)~\emph{Generation} prompts \texttt{GPT-4.1} with the seed, its reranked neighborhood, and a target category to produce an \emph{intra-} or \emph{inter-paragraph} contradiction together with its insertion region (\appref{app:prompts}).
(4)~\emph{Validation} accepts a contradiction only if it is both semantically valid and fluent.

\myparagraph{Validation criteria.}
Contradiction consistency is scored by the NLI model \texttt{nli-deberta-v3-base} as \(\gamma(s,c)=P_{\mathrm{NLI}}(\text{contradiction}\mid s,c)\), and linguistic plausibility by the per-word perplexity shift \(\Delta\mathrm{PPL}_{\text{norm}} = (\mathrm{PPL}(p^{*})-\mathrm{PPL}(p)) / |c|\) under GPT-2, where \(p\) and \(p^{*}\) are the paragraph before and after inserting the contradiction \(c\), and \(|c|\) is the number of words in \(c\).
A contradiction is kept only if \(\gamma \geq \alpha\) and \(\Delta\mathrm{PPL}_{\text{norm}} < \delta\); accepted samples are additionally reviewed manually.

\subsection{Graph-Conditioned Detection Details}
\label{app:detection}

During detection, the input is the contradicted contract \(\mathcal{C}^{*}\) with contradiction locations unknown; the typed paragraph graph \(\mathcal{G}^{*}\) is reconstructed in real time (\appref{app:graph-construction}) and serialized into a structured payload of paragraph nodes with their referential and semantic relations.
This payload is passed to the LLM in a single inference step, enabling document-level reasoning over interconnected graph neighborhoods rather than isolated paragraph pairs.

The detector returns structured predictions following the output schema in \appref{app:prompts}; the parsed fields drive the coordinated views: evidence spans are highlighted in the \textit{Document View}, involved paragraphs are surfaced in the \textit{Related Paragraph Explorer}, and category, evidence, and justification populate the \textit{Contradiction Analysis} panel.
Importantly, the detector is blind to insertion positions: only the experimenters know where contradictions were injected, so performance reflects true discovery rather than position-aware matching.
\autoref{fig:systemreal} shows the \systemname prototype interface.

\myparagraph{Computational cost.}
Typed graph construction remained lightweight (2.3--4.7\,s for contracts of 208--517 paragraphs), while graph-conditioned detection with \texttt{GPT-4.1} cost between \$0.21 and \$0.82 per contract. Larger graph neighborhoods nevertheless increase inference cost and latency for long contracts.
\subsection{Prompt Templates and Output Schema}
\label{app:prompts}

The full verbatim prompts for all LLM-driven steps are released with the source code; we summarize their structure and the detection output schema here.

\myparagraph{Contradiction detection.}
A fixed system prompt configures the model as a legal contradiction classifier with a high-recall bias and requires evidence snippets to be exact verbatim substrings of the contract. The user prompt provides the serialized graph payload---the target paragraph and its typed \texttt{reference}/\texttt{semantic} neighbors---and instructs an intra- and an inter-paragraph pass. Responses follow the schema:

\begin{lstlisting}[language=]
{
  "paragraph_results": [
    {
      "paragraph_id": "string or integer",
      "contradictions": [
        {
          "confidence": integer from 0 to 100,
          "contradiction_type": "temporal | numerical | authority |
                                 process | policy_reversal | specificity",
          "brief_reason": "one short sentence",
          "evidence": {
            "snippet_a": "exact excerpt #1", "snippet_b": "exact excerpt #2",
            "source_a": "paragraph | context | unknown",
            "source_b": "paragraph | context | unknown"
          }
        }
      ]
    }
  ]
}
\end{lstlisting}

\myparagraph{Synthetic contradiction generation.}
The generation prompt embeds the taxonomy of \appref{app:taxonomy} (a definition and example per category) and instructs the model to select a clause, produce a contradictory sentence for insertion into the target paragraph, and label its type and scope, returning \{\texttt{statement}, \texttt{contradiction}, \texttt{type\_contradiction}, \texttt{scope\_contradiction}\}.
\subsection{User Evaluation Questionnaire}
\label{app:user-questionnaire}

After the exploration tasks, participants answered the following items on a five-point Likert scale. \autoref{tab:likert-results} reports the individual responses.

\myparagraph{Usability Items}
\textit{``Was the interface easy to learn?''} (US1);
\textit{``Was the navigation between paragraphs, relationship structure, and contradiction analysis panel clear?''} (US2);
\textit{``Was the presented evidence easy to interpret?''} (US3).

\myparagraph{Usefulness Items}
\textit{``Is the tool useful for supporting real contract review?''} (UF1);
\textit{``Did the contradiction explanations help with my legal validation?''} (UF2);
\textit{``Would I use \systemname in my workflow?''} (UF3).

\userevaluation

\end{document}